\documentclass{article}
\usepackage{amsmath}
\usepackage{graphicx}
\def\p{\partial}

\def\G{\Gamma}
\def\g{\gamma}

\def\d{\delta}
\def\de{\delta}

\def\De{\Delta}
\def\ov{\overline}
\def\ld{\lambda}

\def\ep{\epsilon}

\def\e{\eta}

\def\om{\omega}

\def\b{\beta}

\def\a{\alpha}

\def\pdellx'{\frac{\partial}{\partial x'}}
\def\pdellw'{\frac{\partial}{\partial w'}}

\newcommand{\be}{\begin{equation}}
\newcommand{\ee}{\end{equation}}
\def\bed{\begin{displaymath}}
\def\eed{\end{displaymath}}
\def\bea{\begin{eqnarray}}
\def\eea{\end{eqncrray}}
\def\[{$$}
\def\]{$$}
\begin{document}
\title{Confining Quark Model with General Yang-Mills Symmetry and Inadequate Faddeev-Popov Ghost}

\author{ Jong-Ping Hsu\\
Department of Physics,\\
 University of Massachusetts Dartmouth \\
 North Dartmouth, MA 02747-2300, USA\\
E-mail: jhsu@umassd.edu}
\maketitle
{\small  A quark model with general Yang-Mills $SU_3$ symmetry leads to fourth-order field equations and linear confining potential.  The confining gauge bosons (`confions') are treated as off-mass-shell particles and their indefinite energies are unobservable due to confinement. The ultraviolet divergence of the model appears to be no worse than that of QCD by power counting.   Explicit calculations of the confion self-energy show that the usual Faddeev-Popov ghosts are inadequate to restore gauge symmetry for gauge invariant Lagrangians with higher order derivatives.  `Computer experiments' with FeynCalc lead to a simple empirical method to restore the gauge invariance of the second-order confion self-energy with arbitrary gauge parameters.  The approximate results of third-order vertex corrections suggest that the confining model could be asymptotically free.  
\bigskip


\bigskip



 A unified model for quark confinement and accelerated cosmic expansion with linear potentials was discussed on the basis of  a general Yang-Mills (gYM) symmetry with $(SU_3)_{color} \times (U_1)_{baryon}$.\cite{1,2}  Arbitrary (Lorentz) scalar gauge functions and the phases in usual gauge transformations of fermions such as quarks $q(x)$ are generalized to vector gauge functions $\om_\mu^a (x)$ and a new characteristic function $P(x)$, e.g.,\cite{3,4}
\be
q'(x) = (1- iP)q(x), \ \ \ \   \ov{q}'(x) = \ov{q}(x) (1+iP),   \ \ \ \  P=P(\om,x),
\ee
\be
  P=L^b\left( g_s \int_{x'_o}^x dx'^\mu \om_\mu^b (x') \right)_{Le} \equiv g_s L^b P^b(\om,x), \ \ \  [L^a, L^b]= if^{abc} L^c,
\ee
where $\om\equiv \om^b_\mu(x)$ are infinitesimal arbitrary vector functions (for simplicity) and $\ld^b=2L^b$ are Gell-Mann matrices.\cite{5}     The characteristic phase P is an action integral, which involves a fixed initial point and a variable end point.\cite{4}  We stress that this new phase function P involves an infinitesimal vector gauge function $\om_\mu^a (x)$ and, hence, differs from the usual phase function (which involves a scalar gauge function) in the conventional gauge theories. This is crucial for general Yang-Mills symmetry.  For this new phase function to have unambiguous partial derivatives, we must impose a Lagrange equation (Le) to specify the path, similar to Hamilton's characteristic function,\cite{3} which is a local function and, hence, compatible with local gauge theory. 

 The gYM transformations for the color $SU_{3}$ confining gauge fields $H^a_\mu(x)$ are given by
\be
H'_\mu(x) =H_\mu(x) + \om_\mu(x)  -i[P(x), H_{\mu}(x)], \ \ \ \   H_\mu =H_{\mu}^a L^a,
\ee
where $c=\hbar=1$ and $\e_{\mu\nu}=(+,-,-,-)$.
To see the equation  $Le$ in (2), let us consider the variation of $P$.\cite{4}  We have
\be
 \de P =g_s  \frac{\p \ov{L}}{\p\dot{x}^\ld}\de x^{\ld}  + g_s \left( \int_{\tau_o}^{\tau}\left(-\frac{d}{d\tau}\frac{\p \ov{L}}{\p \dot{x}^\ld}  
 + \frac{\p \ov{L}}{\p x^\ld}\right)\de x^\ld d\tau\right)_{Le},
\ee
where we write (2) in the usual form of a Lagrangian with the help of a parameter $\tau$,
\be
P=\left( g_s \int_{\tau_o}^\tau \ov{L}  \ d\tau\right)_{Le}, \ \ \    \ov{L}=\dot{x}^\mu \om_\mu^a (x) L^a, \ \ \  \dot{x}^\mu=\frac{dx^\mu}{d\tau}.
\ee
We require the paths in (4) to be those that satisfy the Lagrange equation $Le$, i.e., 
\be
-\frac{d}{d\tau}\frac{\p \ov{L}}{\p \dot{x}^\ld} + \frac{\p \ov{L}}{\p x^\ld}=0,    \ \ \  \ov{L}=\dot{x}^\mu \om_\mu^a (x) L^a.
\ee
Thus, the integral in (4) vanishes.  This property leads to an unambiguous relation\cite{4,6}
\be
 \p_{\mu} P^a=\om^a_\mu,  
\ee
which is necessary for the general Yang-Mills symmetry. 

 As usual, the color $SU_{3} $ gauge covariant derivatives are  defined as
\be
 \De_{\mu} = \p_{\mu} + ig_{s}H_{\mu}^{ a} L^a.  
\ee
The $SU_{3}$ gauge curvatures $H^a_{\mu\nu}$ are given by
\be
[\De_\mu, \De_\nu]= ig_s H_{\mu\nu},  \ \ \ \   H_{\mu\nu}=H_{\mu\nu}^{ a} L^a,
\ee
\be
H_{\mu\nu}=\p_\mu H_\nu - \p_\nu H_\mu +i g_s [H_\mu, H_\nu].
\ee

It follows from equations (1)-(10) that we have the following gYM transformations for $\p^\mu H_{\mu\nu}(x)$, and $\ov{q}\De_{\mu} q$:
\be
\p^\mu H'_{\mu\nu}(x)= \p^\mu H_{\mu\nu}(x)- i[P(x), \p^\mu H_{\mu\nu}(x)]
\ee
\be
 \ov{q}' \g^\mu \De'_{\mu} q'  =  \ov{q}\g^\mu  \De_{\mu} q  , 
 \ee
provided the restrictions  
\be
\p^\mu \{\p_\mu \om_\nu (x) - \p_\nu \om_\mu (x)\} - ig_s [\om^\mu (x), H_{\mu\nu}(x)] = 0
\ee
are imposed for (11) to hold.  Nevertheless, we still have infinitely many vector gauge functions $\om^a_\mu (x)$.   This constraint is similar to that for gauge functions of Lie groups in the usual non-Abelian gauge theories.\cite{7}
The general Yang-Mills transformations have group properties and reduce to usual gauge transformations in special cases.\cite{6} 
  
Let us concentrate on the $SU_3$ sector of the gYM invariant Lagrangian, which is assumed to be
\be
L_{gYM} = \frac{L_s^2}{2}[ \p^\mu H^a_{\mu\ld} \p_\nu H^{a\nu\ld}]
+i \ov{q}(x)\g^\mu\De_{\mu} q (x)- m_q \ov{q}(x)q(x) . 
\ee
It leads to  the fourth-order gYM field equation,
\be
\p^2 \p^\mu H^a_{\mu\nu} - (g_s/L^2_s)\ov{q} \g_\nu (\ld^a/2) q = 0.
\ee
In the static case, one has the source terms, $ -g_s \de^3({\bf r}) + g_s L_s^2\nabla^2 \de^3({\bf r})$, where the first term is due to the usual point source of quark, while the second term could be generated from the self-coupling of the confion fields.\cite{6}  These sources produce the dual static potentials,
\be
H_0 =  g_s [ r/(8\pi L_s^2) - 1/(4\pi r)], \ \ \ \ \  \frac{g^2_s}{4\pi} \approx 0.04, \ \ \ \ \ \ \  L_s \approx 0.28 fm,
\ee
which provides a  mechanism for quark confinement.  They are just right to support the ideas and results of Cornell group,\cite{8} which enables us to determine the values  of $L_s$ and the confion coupling strength $g_s^2/4\pi$ in (16).
 The length $L_s$ denotes a universal length and could play a role in particle-cosmology.\cite{6,9}  In (16), $L_s$ could be considered as the universal and fundamental length for all gYM gauge fields. The coupling strength ${g^2_s}/{4\pi} \approx 0.04$ suggest that one could do reliable perturbative calculations in the confining model.  The cosmic implications of the gYM fields (with $(SU_3)_{color} \times (U_1)_{baryon}$) for the late-time cosmic acceleration was discussed in previous works.\cite{6,9}  

 The fourth-order field equation is usually considered unphysical because the dynamical system involves non-definite energy, otherwise there is no essential difficulty, according to Pais and Uhlenbeck, and others.\cite{10,11,12}  However, in the present confining model with general Yang-Mills symmetry, the massless confions, which satisfy the fourth-order equation (15), are permanently confined in the quark system and, hence, their negative energies cannot be detected with the present apparatus.   The confining model suggests that these new gauge bosons could be treated as `off-mass-shell particles' in the external states and intermediate states of the S matrix and, hence, do not contribute to the imaginary part of physical amplitudes to violate unitarity.\cite{13}  Some physicists hoped that higher-order fields might help to eliminate ultraviolet divergences and to construct a finite quantum field theory.  But it is very difficult to formulate a finite field theory without having the problems of indefinite energies and non-unitarity of the S matrix.  
 We note that previous investigations of higher-order Lagrangians did not include gauge fields with dynamical $SU_N$ symmetry groups.  Indeed, the situation changes when the ideas of gYM symmetry and confinement are introduced simultaneously.  It appears that for the gYM symmetry with simple $(U_1)_{baryon}$ group, one could have a field theory with a finite fermion self-energy.  When this is applied to neutrinos, the result may have interesting implications to neutrino oscillations and dark matter because neutrinos must have non-vanishing masses.\cite{14}

 There are non-trivial and qualitative differences between the propagations of waves satisfying fourth-order and second-order equations.  Nevertheless, we shall assume that the formulation of local field theory and Dyson's derivation of the rules for Feynman diagrams could be applied to the new Lagrangian (14) with higher order derivatives, including Faddeev-Popov (FP) ghost fields.  
 
 For the following discussions of Feynman-Dyson rules, confion self-energy and one-loop corrections to the 3-confion vertex, let us concentrated on the pure confion part in the Lagrangian (14) for simplicity.
To obtain the Feynman-Dyson rules in the confining model, the vacuum-to-vacuum amplitude (i.e., the generating functional for connected Green's functional) for the confion field is assumed to take the conventional form,\cite{15,16}
\be
W_{Y_c}(j_c)=\int d[H^a_{\mu}]  exp\left(i\int d^4 x (L_{gYM}(H) + H^a_{\mu}j_c^{a\mu} )  \frac{}{} \right) 
\ee
$$
 \times \  det U_c \   \mbox{\large \boldmath $ \Pi$}_{x, a}\left[\frac{}{} \de(\p^2 \p_{\mu} H^{a\mu} - Y_{c}^a)\right] .
$$
The  gauge condition, $\p^2 \p_{\mu} H^{a\mu} = Y_c^a$, for the gYM invariant Lagrangian is assumed so that the corresponding ghost fields also satisfy the fourth-order equations, similar to the confion fields.  Its presence in the delta functions in (17) may be considered as to force the vacuum-to-vacuum amplitude (17) to maintain the same gauge conditions for all time and to restore the gYM symmetry of the S matrix.\cite{13}  Whether this new gauge condition works as expected in the confining model will be examined below by explicit calculations.   
As usual, we may write $W[j_c]$ as\cite{16}
$$
W[j_c]= \int d[Y_c] W_{Y_c}(j_c) exp\left[ -i\int d^4x L_s^2\left(\frac{1}{2\xi} Y^a_c \p^{-2} Y^a_c \right)\right]
$$
$$
=\int d[H^a_{\mu}]  exp\left(i\int d^4 x (L_{gYM}+ H^a_{\mu}j_c^{a\mu})\right) detU_c 
$$
\be
=\int d[H^a_{\mu},D^b,\ov{D}^c]  exp\left(i\int d^4 x (L_{gYM}(H) + L_{gf} + L_{ghost}+ L_j)\right);
\ee
$$
L_{gYM}(H)+L_{gf} =\frac{L_s^2}{2} \p^\mu H^a_{\mu\ld} \p_\nu H^{a\nu\ld}+ \frac{L_s^2}{2\xi}(\p^{\ld}\p_{\mu}H^{a\mu})(\p_{\ld}\p_{\nu}H^{a\nu}),   
 $$
 $$
L_{ghost}=L^2_s  \p^2 \ov{D}^{a}\left[\p^2 D^{a} -g_s f^{abc} \p^\mu(H^b_\mu D^c)\right], \ \ \  L_j= H^a_{\mu}j_c^{a\mu},
 $$
where $\p^{-2}$ is an integral operator, and $\xi$ is an arbitrary gauge parameters

The Feynman-Dyson rules for Feynman diagrams can be obtained from the effective Lagrangian $L_{eff}=L_{gYM}(H) + L_{gf} + L_{ghost}$ in (18), where the scalar fermion field $\ov{D}^i$ is considered to be independent of $D^i$.  
 By definition of the physical subspace of 
states for the S-matrix, these ghost particles do not appear in the 
external states.  They can only appear in the intermediate 
states of a physical process.\cite{13} 

The confion propagator can be obtained from their  fourth-order field equations (15),  
\be
C^{ab}_{\mu\nu}(k) =\frac {-i\de^{ab}}{L^2_s (k^2+i\ep)^2}\left[\e_{\mu\nu} -(1-\xi)\frac{k_\mu k_\nu}{k^2+i\ep}\right], 
 \ee
The Faddeev-Popov (FP) ghost propagator for $D_a(x)$ is given by $L_{ghost}$ in (18),
\be
G_{ab}= \frac{-i\de_{ab}}{L_s^2(k^2+i\ep)^2}, 
\ee
The 3-confion vertex $[H^a_\a(k_1)H^b_{\b}(k_2)H^c_{\g}(k_3)]$ is given by, 
$$
g_s L^2_s f^{abc}[(k_1)^2( k_{2\b}\e_{\a\g}- k_{3\g}\e_{\a\b} + k_{3\b}\e_{\a\g} - k_{2\g}\e_{\a\b})
$$
$$
+ (k_2)^2( k_{3\g}\e_{\a\b}- k_{1\a}\e_{\g\b} + k_{1\g}\e_{\a\b} - k_{3\a}\e_{\g\b})
$$
\be
+ (k_3)^2( k_{1\a}\e_{\b\g}- k_{2\b}\e_{\a\g} + k_{2\a}\e_{\b\g} - k_{1\b}\e_{\a\g})],
 \ee
and the ghost vertex $[\ov{D}^a(k_1) D^b(k_2)H^c_\mu(k_3)]$ takes the form
\be
  -g_s L^2_s f^{abc} (k_1\cdot k_1)k_{1\mu}, 
\ee 
where $(k_n)^2=k_{n\mu}k_n^\mu, \  n=1,2,3$, and all momenta in (21) and (22) flow into the vertex.  The 4-confion vertex (which can be combined into 72 terms) can also be obtained from $L_{gYM}(H)$ in (18).  Other rules such as  a factor -1 for each ghost loop, etc. are the same as those in usual gauge theories.
  
Although the model appears to be gauge invariant and renormalizable by power counting,  we would like to check gauge invariant amplitudes by explicit calculations of the confion self-energy.  The reasons are that the model is more complicated and more involved due to the general Yang-Mills symmetry and the presence of fourth-order field equations, as well as  the restriction (13) for the vector gauge functions.   

\bigskip
\noindent
(A) Confion self-energy.
\bigskip
 
 Based on dimensional regularization,  calculations of the confion self-energy could test whether the usual Faddeev-Popov method is inadequate to restore gauge invariance of the S matrix.
With the help of computer programs, FeynCalc and FeynHelpers,\cite{17,18,19} one can evaluates the ultraviolet divergent part of the confion self-energy to see if the general Yang-Mills invariance is satisfied, similar to that of the usual non-Abelian gauge symmetry of QCD. Let us consider the Lagrangian $L_{gYM}(H) + L_{gf}+L_{ghost}$ given by (18).  The free confion field in $L_{gYM}$ can be expressed in the form $(L_s^2/2)H^{a\mu}(\p^2\p^2\e_{\mu\nu}-\p^2 \p_\mu \p_\nu)H^{a\nu}$.  The amplitude for the confion proper self-energy part takes the usual gauge invariant form
\be
 \mbox{\large  $ \Pi$}_{\mu\nu}^{ab}(p)=\de_{ab}(p_\mu p_\nu - p^2 \e_{\mu\nu})  \mbox{\large  $ \Pi$}(p),
\ee
which diverges logarithmically.  There are three diagrams contribute to the one loop process $H^a_\mu(p)\to (one-loop)\to H^b_\nu(-p)$ of the confion self-energy.  In general, the confion propagator (19) involves an arbitrary gauge parmeter $\xi$.  Nevertheless, if one chooses $\xi=0$ initially, it remains zero under finite renormalizations.\cite{20}  For simplicity, we use the Landau gauge, $\xi=0$, and dimensional regularization to evaluate the three divergent Feynman diagrams for the confion self-energy.  We obtain the following results:
 
 (i) One loop with two 3-confion vertices (3c{\em v}): 
 \be
 \mbox{\large  $ \Pi$}_{\mu\nu}^{ab}(p,3cv)=iQ\frac{ \d^{ab}}{4}(7 p_\mu p_\nu -10 p^2 \e_{\mu\nu}), \ \ \ \ \  Q\equiv\left(\frac{g_s^2}{16\pi^2}\right)\frac{C_A}{ \ep_{UV}}.
\ee

(ii) One loop with one 4-confion vertex (4c{\em v}): 
 \be
 \mbox{\large  $ \Pi$}_{\mu\nu}^{ab}(p,4cv)= 0.
\ee

(iii) One FP ghost loop with two confion-ghost vertices ({\em ghost}): 
 \be
 \mbox{\large  $ \Pi$}_{\mu\nu}^{ab}(p,ghost)=iQ\frac{ \d^{ab}}{12}(2 p_\mu p_\nu + p^2 \e_{\mu\nu}),
\ee
where $1/\ep_{UV}$ diverges logarithmically and $C_A=N$ for $SU_N$.

The confion-loop contribution (24) by itself does not satisfy the expected form (23) of the gauge invariant, similar to the corresponding diagram withs the gluon-loop in QCD.  Moreover, the tadpole-like contribution (25) vanishes due to dimensional regularization, the same as what happens for the corresponding gluon tadpole-like diagram in QCD.  When we add the FP-ghost loop contribution (26) to (24), the resultant amplitude does not satisfy the expected gauge invariant form (23):
\be
 \mbox{\large  $ \Pi$}_{\mu\nu}^{ab}(p,3v)+ \mbox{\large  $ \Pi$}_{\mu\nu}^{ab}(p,ghost)=iQ\frac{ \d^{ab}}{12}(23 p_\mu p_\nu -29 p^2 \e_{\mu\nu}).
\ee
  In contrast, the amplitudes corresponding to (24)-(26) in QCD are respectively given by 
$$
\left[iQ\frac{ \d^{ab}}{12}(-28p_\mu p_\nu +25p^2 \e_{\mu\nu})\right], \ \ \ [0], \ \ \  \left[iQ\frac{ \d^{ab}}{12}(2p_\mu p_\nu  +p^2 \e_{\mu\nu})\right]. 
$$
Thus, the sum of contributions from gluons and Faddeev-Popov ghosts in QCD leads to the gauge invariant 
result, $iQ\d^{ab}(13/6)(p^2 \e_{\mu\nu}-p_\mu p_\nu)$.
 
The non-gauge-invariant result (27) reveals a problem in the confining quark model with  gYM symmetry.  We perform more `computer experiments' to explore the possibility of obtaining gauge invariant form (23) based on the amplitudes in (24)-(26).  By a stroke of luck, we find out that the following combination leads to the gauge invariant form (23):
\be
\mbox{\large  $ \Pi$}_{\mu\nu}^{ab}(p,3v)+ 3 \mbox{\large  $ \Pi$}_{\mu\nu}^{ab}(p,ghost)=iQ\frac{9 \d^{ab}}{ 4}( p_\mu p_\nu - p^2 \e_{\mu\nu}), \ \ \ \  \xi=0.
\ee
Furthermore, by doing more `computer experiments,' we also find out that the combination (28) turns out to be `correct' for an arbitrary gauge parameter $\xi$,
\be
\mbox{\large  $ \Pi$}_{\mu\nu}^{ab}(p,3v)+ 3 \mbox{\large  $ \Pi$}_{\mu\nu}^{ab}(p,ghost)=iQ \frac{ \d^{ab}}{4 }(\xi+3)^2( p_\mu p_\nu - p^2 \e_{\mu\nu}). \
\ee
This interesting result for arbitrary gauge parameters suggests that there is a method for introducing a `correct ghost' in the confining model and that it will work for arbitrary gauges.  Moreover, it shows that the one-loop corrections in the confining model depends on $\xi^2$.  In contrast, they depend only on $\xi$ at the one-loop level in QCD.\cite{20,21}   We hope that it could shed some light on further investigation.  

\bigskip
\noindent
 (B) Possible asymptotic freedom of the confining quark model
\bigskip

We explore further the gauge invariant property through more `computer experiments' for one-loop corrections to the 3-confion vertex and for possible asymptotic freedom of the model.  
 Let us consider and estimate the sign of the $\b-$function related to the asymptotic freedom of the confining model based on the Feynman-Dyson rules (19)-(22), etc.  
 We are interested in the ultraviolet (UV) divergent part of the 3-confion vertex function,
\be
 \G^{abc}_{\ld\mu\nu}(0,p,-p)=(2p_{\ld}\e_{\mu\nu}-p_{\mu}\e_{\ld\nu}-p_{\nu}\e_{\ld\mu})f^{abc}g_s F(p^2/M^2, g^2_s),
\ee
with the subtraction at $p^2=M^2$ to avoid the infrared divergent.  In order to simplify the calculations with an arbitrary gauge parameter $\xi$ in the confion propagator (19), we choose specific kinematics, $p_{1\mu}=0$ for incoming confion and $p_{2\mu}=p_\mu=-p_{3\mu}$ for two outgoing confions, as indicated by $ \G^{abc}_{\ld\mu\nu}(0,p,-p)$ in (30). 

 At the one-loop level, we obtain the UV divergent parts of the vertex corrections with an arbitrary gauge parameter $\xi$.  For two diagrams involving FP ghost-triangles (2ghost), we have the usual $\xi$-independent result,
\be
\G_{\ld\mu\nu}^{abc}(2ghost)=g_s Q\frac{ f^{abc}}{24}[2p_{\ld}\e_{\mu\nu}-p_{\mu}\e_{\ld\nu}-p_{\nu}\e_{\ld\mu}].
\ee
 A bubble diagram is a one-loop diagram with a 3-confion vertex and a 4-confion vertex on it.  For three diagrams with confion-bubbles (3cb), we obtain
$$
\G_{\ld\mu\nu}^{abc}(3cb)= - g_s Q\frac{ f^{abc}}{24}[2p_{\ld}\e_{\mu\nu}(6\xi^2+28\xi+44)
$$
\be
 - (p_{\mu}\e_{\ld\nu}+p_{\nu}\e_{\ld\mu})(9\xi^2+34\xi+41)].
\ee
One of the three bubble diagrams having two external confions coming from the same 4-confion vertex on a loop (similar to a tadpole diagram) vanishes due to dimensional regularization.  For the diagram with one confion-triangle (1ct), we have
$$
\G_{\ld\mu\nu}^{abc}(1ct)= - g_s Q\frac{ f^{abc}}{8}[2p_{\ld}\e_{\mu\nu}(12\xi^2+60\xi+100)
$$
\be
- (p_{\mu}\e_{\ld\nu} +p_{\nu}\e_{\ld\mu})(19\xi^2+71\xi+82)].
\ee

From (31) to (33), we see that the sum of all these vertex corrections does not satisfy the form (30) for the vertex, similar to the situation in the confion self-energy.
Note that the FP ghost contribution (31) is $\xi$-independent and satisfies the form (30) already and, hence, cannot help the $\xi$-dependent results in (32) and (33) to satisfy the required form (30).  This result also indicates that the usual Faddeev-Popov method alone is inadequate for the confining quark model based on general Yang-Mills symmetry.
 
Let us consider the possible asymptotic freedom of the confining model.  We note that the sum of the vertex corrections in QCD from the gluon and FP ghost contributions in the Landau gauge is given by:
 \be
 -g_s Q\frac{17 f^{abc}}{12}\left[2p_{\ld}\e_{\mu\nu}-(p_{\mu}\e_{\ld\nu}+p_{\nu}\e_{\ld\mu})\right], \ \ \ \  \xi=0,
 \ee
which has the expected form and, together with the gluon self-energy, leads to the asymptotic freedom of QCD.\cite{20,21} 
 In contrast, the amplitudes corresponding to (34) contributed by confions and ghosts in the confining model are approximately given by 
\be
g_s Q f^{abc}\left[2p_{\ld}\e_{\mu\nu}\left(\frac{-343}{24}\right)-(p_{\mu}\e_{\ld\nu}+p_{\nu}\e_{\ld\mu})\left(\frac{-286}{24}\right)\right], \ \  \ \ \ \ \ \  \xi=0,
\ee
 where ghost contribution (31) is less than 1\%.  

Based on gauge symmetry, it seems reasonable to conjecture that when a 'correct ghost' contributes to modify the vertex correction (35), one will have the approximate result,
\be
g_s Q f^{abc} B \left[2p_{\ld}\e_{\mu\nu}-(p_{\mu}\e_{\ld\nu}+p_{\nu}\e_{\ld\mu})\right], \ \  \  -14\le B \le -12,
\ee
where B could have values between the two coefficients (-343/24) and (-286/24) in (35).  

In the confining quark model with gYM $SU_3$ symmetry, the renormalized Lagrangian $L_{rgYM}$ is defined by
$$
L_{rgYM}=Z_3(L_s^2/2)H^{a\mu}(\p^2\p^2\e_{\mu\nu}-\p^2 \p_\mu \p_\nu)H^{a\nu} 
$$
\be
-Z_{3c}L_s^2 g_s f^{abc}(\p^2 \p_\nu H^a_\ld) H^{b\nu} H^{c\ld}+....,
\ee
which are relevant terms for our discussions of possible asymptotic freedom. 
Based on considerations of self-energy and vertex corrections, the confining model with gYM symmetry lead to the approximate results
\be
Z_3\approx 1+Q\frac{9}{2 L_s^2 p^2}, \ \ \  Z_{3c}\approx 1+ Q B,
\ee
where we have included the counter-terms contributions in the computations of the confion self-energy and 3-confion vertex.  The presence of the factor $L_s^2 p^2$ is due to the fourth-order gauge field equations and the basic length scale $L_s$ is given by (16).   Based on considerations of gauge symmetry and the approximations (29), (36) and (38), the sign of the $\b$ function may be determined approximately by
\be
\b(g_s) \approx \left(\frac{g_s^3}{16\pi^2}\right)\frac{C_A(-27+4B)}{2 L_s^2 p^2}.
\ee
which is negative.  It suggests that the confining quark model could be asymptotically free, provided the value of B is approximately given by (36). We have also considered the Feynman gauge, the situation is similar to that of the Landau gauge:  There is a corresponding parameter $B_1$, which may take the values roughly between -14 and -25, and hence the $\b$ function will also be negative. The result (39) is not unreasonable because the confining model is also based on gYM symmetry with $SU_3$ and the contributions from the ghost particles appear to be small in general.

The usual linear gauge conditions such as $\p_\mu H_a^\mu = y_a(x)$ or $\p^2 \p_{\mu} H^{a\mu} = Y_c^a$ in (17) imply that the corresponding FP ghost can only contribute $\xi$-independent amplitudes at the one-loop level.  Only if one chooses a non-linear gauge condition, one can have an addition parameter in the ghost interaction without violating unitarity of the theory.\cite{13}  But (31)-(33) show that the usual Faddeev-Popov method with the gauge condition in (17) is inadequate and that the `correct ghost' must be able to contribute $\xi$-dependent amplitudes to restore gauge invariance at the one-loop level.  In this aspect, perhaps, the `correct ghost' should resemble the Feynman-DeWitt-Mandelstam ghost in quantum Yang-Mills gravity.\cite{16}  The reason is that if the ghosts are massless `vector fermions,' their propagators could involve a gauge parameter $\xi$.  Moreover, the restriction (13) for the vector gauge function $\om_\mu(x)$ could affect the interaction of the correct ghost.\footnote{In this case, it may be more convenient to use Lagrangian multiplier method to handle the restriction (13).\cite{22} } Presumably, all these issues and the $\xi^2$-dependence of amplitudes in (32)-(33) could make the solution to the problem non-trivial.  

However, the observable results calculated with perturbation in the confining model are reliable because of the small confion coupling strength $g_s^2/4\pi\approx 0.04$ in (16), without appeal to the asymptotic freedom at high energies.  Furthermore, the 3-confion vertex and amplitudes with external confions are not observable because of the permanent confinement of confions in quark systems by linear potentials.  In light of previous discussions, the confining quark model appears to be interesting and deserves further investigations.

 The work was partially supported by the Jing Shin Research Fund of the UMassD Foundation.  The author would like to thank Y. Hao for assistance. He would  also like to thank V. Shtabovenko and R. Mertig  for help and for providing a `super-Feynman-toolbox' to shed light on complicated one-loop amplitudes and to do `computer experiments.' 


\end{document}